\documentclass{article}
\usepackage[margin=1.25in]{geometry}
\usepackage{parskip}
\usepackage{amsmath}
\usepackage{graphicx}
\usepackage[hidelinks]{hyperref}
\usepackage{natbib}
\begin{document}
\title{Selecting velocity models using Bayesian Information Criterion}
\author{Tomasz Danek\footnotemark[1], Bartosz Gierlach\footnotemark[1], Ayiaz Kaderali\footnotemark[2], Michael A. Slawinski\footnotemark[2], \\ and Theodore Stanoev\footnotemark[2]}
\date{}
\footnotetext[1]{
AGH University of Science and Technology, 
Department of Geoinformatics  and Applied Computer Science, Krakow, Poland. E-mail: tdanek@agh.edu.pl, gierlach@agh.edu.pl}
\footnotetext[2]{Memorial University of Newfoundland, 
Department of Earth Sciences, 
St. John's, Canada. E-mail: ayiazkaderali@gmail.com, mslawins@mac.com, theodore.stanoev@gmail.com}
\maketitle
\begin{abstract}
We present a strategy for selecting the values of elasticity parameters by comparing walk-away vertical seismic profiling data with a multilayered model in the context of Bayesian Information Criterion. 
We consider $P$-wave traveltimes and assume elliptical velocity dependence. 
The Bayesian Information Criterion approach requires two steps of optimization. 
In the first step, we find the signal trajectory and, in the second step, we find media parameters by minimizing the misfit between the model and data. 
\end{abstract}
\section{Introduction}
In this paper, we use Bayesian Information Criterion (BIC) to select a justifiable parameterization of a model \cite[]{schwarz}.
To restrict the parameterization from an infinity of models, we use explicit and implicit selection criteria.
For the former, as traveltime inversion is one of the most important techniques for extracting information on the Earth's properties~\citep[e.g.,][ Section~9.4]{AkiRichards2002}, we select four traveltime parameterizations to be considered in BIC.
For the latter, we impose a range of elasticity-parameter values that are consistent with sedimentary basins.
We use ray theory, and assume elliptical velocity dependence, to solve a~two-stage optimization problem to obtain the elasticity parameters of a~multilayer medium.
The dataset contains traveltimes for a wide range of offsets, which is necessary to examine anisotropy.

Quantitative analysis of seismic wave propagation is essential in seismic interpretation. 
Such an analysis is complicated even for relatively simple cases, such as horizontally layered media. 
Ray theory, which is invoked in this work, provides mathematical tools that simplify the analysis~(e.g., \cite{keller,cerveny,shearer,slaw99,wang,slaw03,slaw04}).
\section{Theory}
\subsection{Elliptical velocity dependence}
We obtain signal traveltime in an anisotropic inhomogeneous medium by considering stationary traveltimes within a given velocity model. 
We consider inhomogeneity, $V(z)=a+bz$, where $a$ and $b$ are constant, $z$ is the vertical component that corresponds to depth, and anisotropy~\citep{slaw04}
\begin{equation}
	\label{eq:chi}
	\chi = \frac{v_h^2-v_v^2}{2 v_v^2}\,,
\end{equation}
where $v_v$ and $v_h$ are the vertical and horizontal speeds.
Expression~\eqref{eq:chi} describes an elliptical velocity dependence of a wavefront.
For $v_h = v_v$, $\chi = 0$ and, hence, the wavefront velocity is isotropic.

For a source placed at point $(0,0)$ and receiver at $(x,z)$ the traveltime is \cite[]{rogister}

\begin{equation} \label{eq:t}
	t=\frac{1}{b} \left\lbrace \textrm{arctanh}\left[ pbx - \sqrt{1-(1+2\chi)p^2a^2}\right]+\textrm{arctanh}{\sqrt{1-(1+2\chi)p^2a^2}}    \right\rbrace\,,
\end{equation}

where

\begin{equation}
	p=\frac{2x}{\sqrt{\left[{x^2+(1+2\chi)z^2}^{\rule{0mm}{1.5mm}}\,\right]\left[{(2a+bz)^2(1+2\chi)+b^2x^2}^{\rule{0mm}{1.5mm}}\,\right]^{\rule{0mm}{1mm}}\,}}
\end{equation}

is the ray parameter, which is a conserved quantity along the ray.
In keeping with SI units, the units for $a$ and $b$ are m/s and 1/s, respectively, for speed are m/s, for traveltime are s\,, and for the ray parameter are s/m.
\subsection{Ray optimization in multilayered media}
We consider the aforementioned model with layer interfaces based on VSP measurements \citep{ayiaz}.
Each layer is characterized by the values of $a,b,\chi$\,.
In each layer, the traveltime along a~ray is given by expression~\eqref{eq:t}.
We consider a two-step optimization.
First, the signal trajectory is optimized for each source-receiver pair to obey Fermat's principle, for a set of the $a,b,\chi$ values.
Second, these values are adjusted to minimize the misfit between the modelled and measured traveltimes.
These steps are repeated until the misfit is at a minimum; the misfit is used in the BIC context.

Both steps are performed using the Nelder-Mead simplex method, which is a local optimization.
Since it is not based on the gradient, it can be used for nondifferentiable functions. 
The method works for functions of $n$ variables, whose values are calculated at $n+1$ points in an $n$-dimensional solution space. 
These points are the vertices of a~polyhedron called a~simplex. 
Successive steps of optimization consist of adjusting these vertices according to specific rules. 
The detailed description of the method can be found in \cite{nelder}.
\subsection{Bayesian Information Criterion}
The optimization requires setting the number of parameters {\it a priori}.
This number should be chosen to match the resolving power of the data. 
For that purpose, we use BIC, whose most general form is~\citep[ equation~23]{kass}
\begin{equation}
	BIC = -2\ln L + k \ln M \,,
\end{equation}
where $L$ is maximized likelihood, $k$ is the number of model parameters, and $M$ is the number of data points, which, herein, is the number of traveltimes. 
According to \citet[ p.~375--376]{priestley}, the same minimum value is obtained by minimizing
\begin{equation} \label{eq:bic}
	BIC = M \ln \hat{\sigma} ^2 +k\ln M  \,,
\end{equation}
where $\hat{\sigma}^2$ is the error variance, which, herein, is the normalized mean of squared differences between measured and modelled traveltimes.
The model whose BIC value is the least is considered best in terms of balance between agreement with measurements and model
complexity.
Compared to other criteria, such as Akaike Information Criterion, BIC results in a~bigger penalization for additional parameters \cite[]{kass}. 
The BIC method is commonly used in similar studies~(e.g.,~\cite{bic-mt}, \cite{danek}).
\section{Results}
\subsection{Data and initial models}
The dataset used in this paper consists of VSP measurements from offshore Newfoundland~\cite[]{ayiaz}.
The walk-away VSP data is the basis for the inversion; the zero-offset VSP is used to get the initial model.

For the zero-offset VSP, receivers are in the entire well at 30\,m intervals.
For the walk-away VSP, there are two-hundred source locations with 25\,m intervals along the NW-SE line. 
Maximum offset is 4000\,m, in the NW direction, and 1000\,m, in the SE direction. 
The receiver array consists of five geophones at depths between 1980\,m and 2020\,m, with respect to mean sea level. 
In accordance with \cite{ayiaz}, the near-offset data, up to 300\,m, are removed to avoid problems with solution stability for near-vertical rays. 
For both types of VSP data we consider $P$-wave traveltimes only.
 
\begin{figure}
\centering
\includegraphics[scale=0.5]{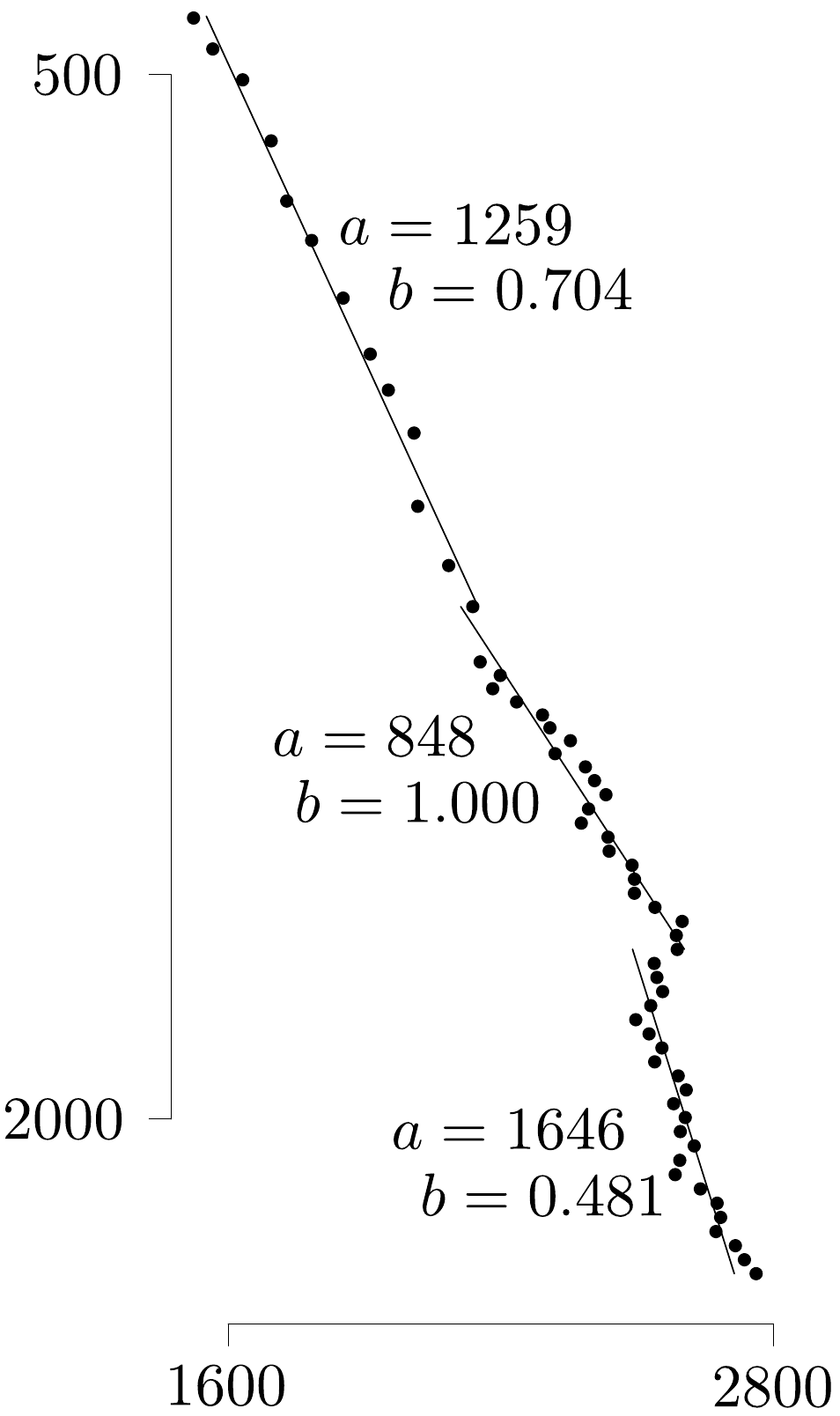}
\caption{\small
	Interval velocities based on zero-offset VSP data used to determine the starting three-layer model for inversion.
	The units for $a$ are~m/s and for $b$ are~1/s.
	Depth along the vertical axis is in~m whereas velocities along the horizontal axis are in~m/s.
}
\label{fig:vel_curve}
\end{figure}

Figure~\ref{fig:vel_curve} is obtained by smoothing the VSP data by exponential smoothing.
We observe three distinct velocity gradients, thus, we infer a three-layer model, whose interfaces are at 1300\,m and 1750\,m, as the initial model used in all computations.

Our simplest model is inhomogeneous, but isotropic, and consists of six parameters $a_i$, $b_i$, where $i=1,2,3$. 
The most complicated model is inhomogeneous and anisotropic, and consists of nine parameters $a_i$, $b_i$, and $\chi_i$, where $i=1,2,3$. 
In between these two extremes, we consider a seven- and eight-parameter model, wherein the middle and first two layers are anisotropic, respectively.
A similar analysis---for synthetic data---is described by \cite{gierlach}.
\subsection{Inversion models}
\begin{figure}
\centering
\includegraphics[scale=0.5]{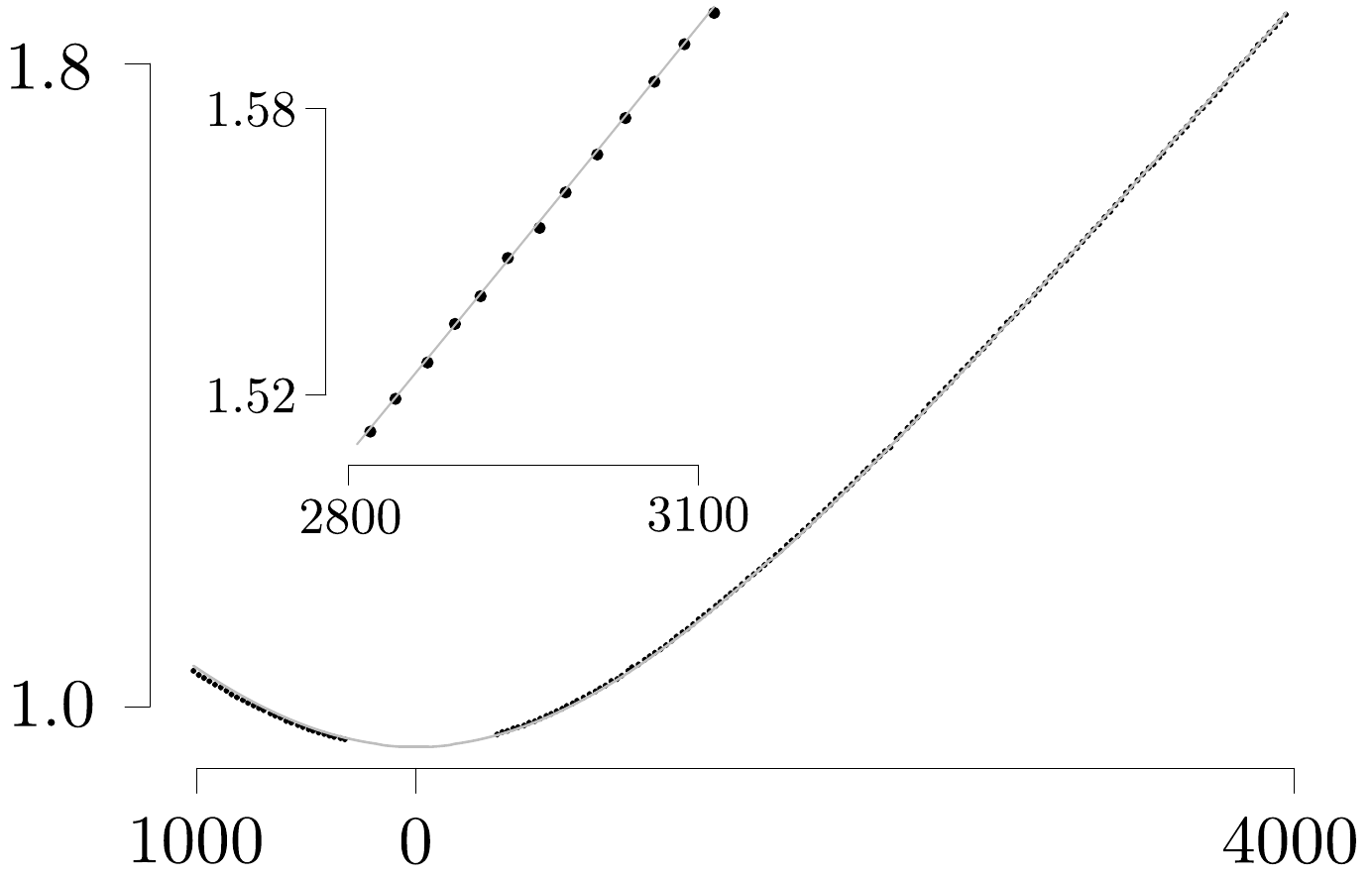}
\caption{\small
	Comparison of obtained (line) and measured (points) traveltimes for receiver at depth of 2020\,m using the seven-parameter model; details illustrated within the insert.
	Traveltimes along the vertical axis are in~s whereas offsets along the horizontal axis are in~m\,.
}
\label{fig:czasy}
\end{figure}

We obtain consistent results for each of the models regardless of their complexities.
For example, in Figure~\ref{fig:czasy}, we see the results for a seven parameter model.

Since the chosen optimization method is local, the parameters obtained depend on the initial model. 
To obtain consistent results, we use a multistart procedure for a~wide range of initial values. 
In other words, the inversion is performed numerous times with randomly chosen initial-model values.
The final results correspond to the least misfit. 
Also, we used the multistart analysis to examine interdependences between parameters, as illustrated in Figure~\ref{fig:a0b0,a1chi1}.

The multistart analysis diminishes the dependence of results on the initial model. 
Thus, values obtained can be treated as global extrema. 
A reasonable initial model is a~set of values obtained from the zero-offset VSP with the addition of small elliptical anisotropy for the middle layer. 

\begin{figure}
	\centering
	\includegraphics[scale=0.5]{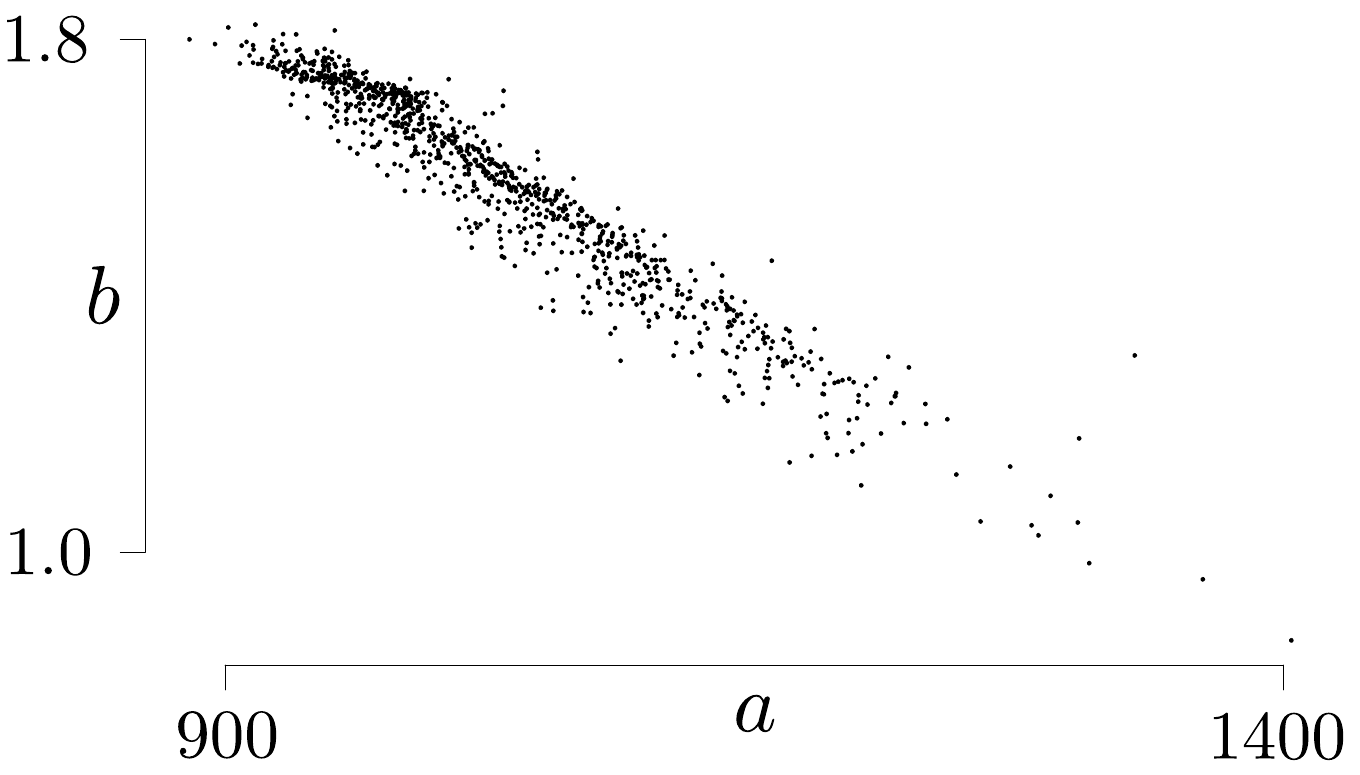}
	\quad
	\includegraphics[scale=0.5]{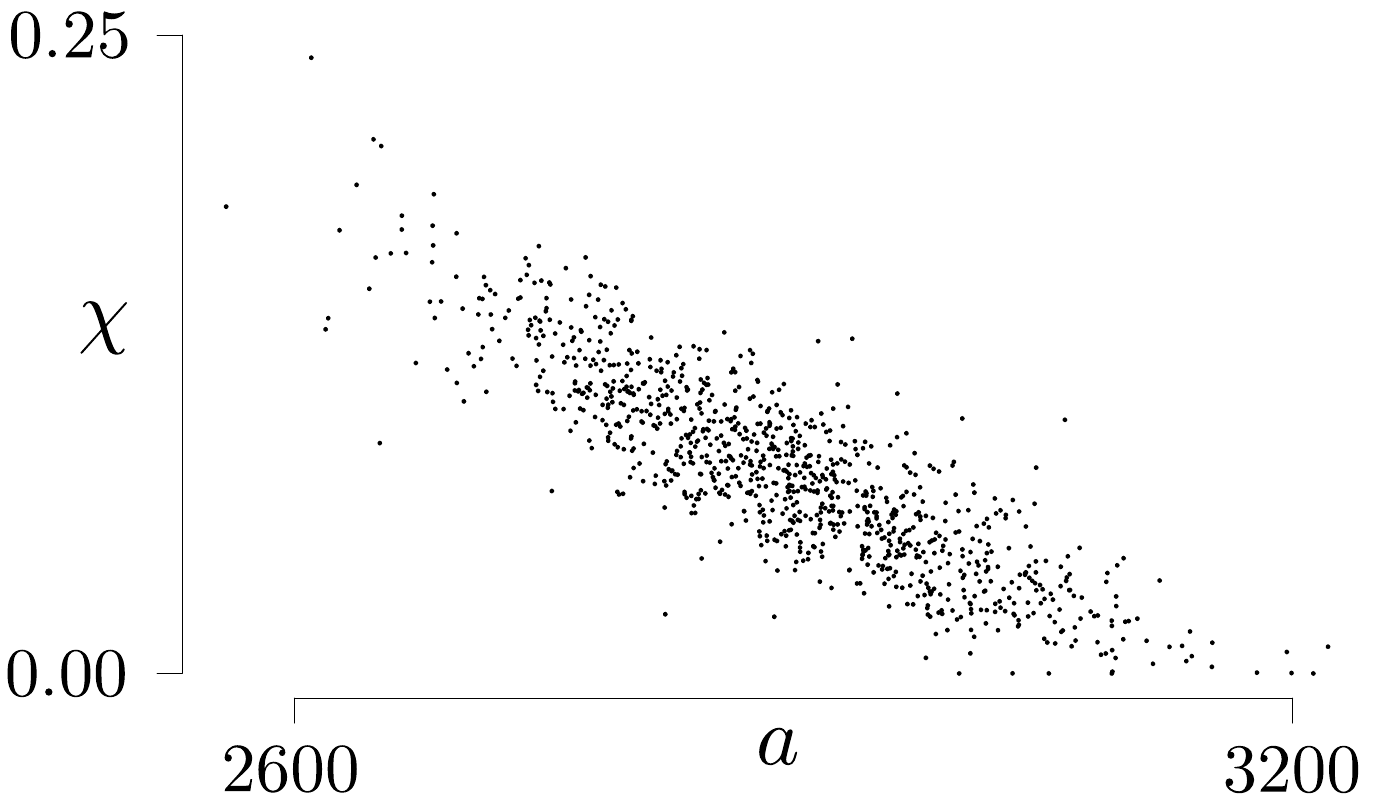}
	\caption{
		Crossplots of values for parameters obtained for a seven-parameter model.
		The left- and right-hand plots correspond to the first and second layers, respectively.
		The units of $a$\,, $b$\,, and $\chi$ are m/s, 1/s, and unitless.
	}
	\label{fig:a0b0,a1chi1}
\end{figure}

\begin{figure}
\centering
\includegraphics[scale=0.5]{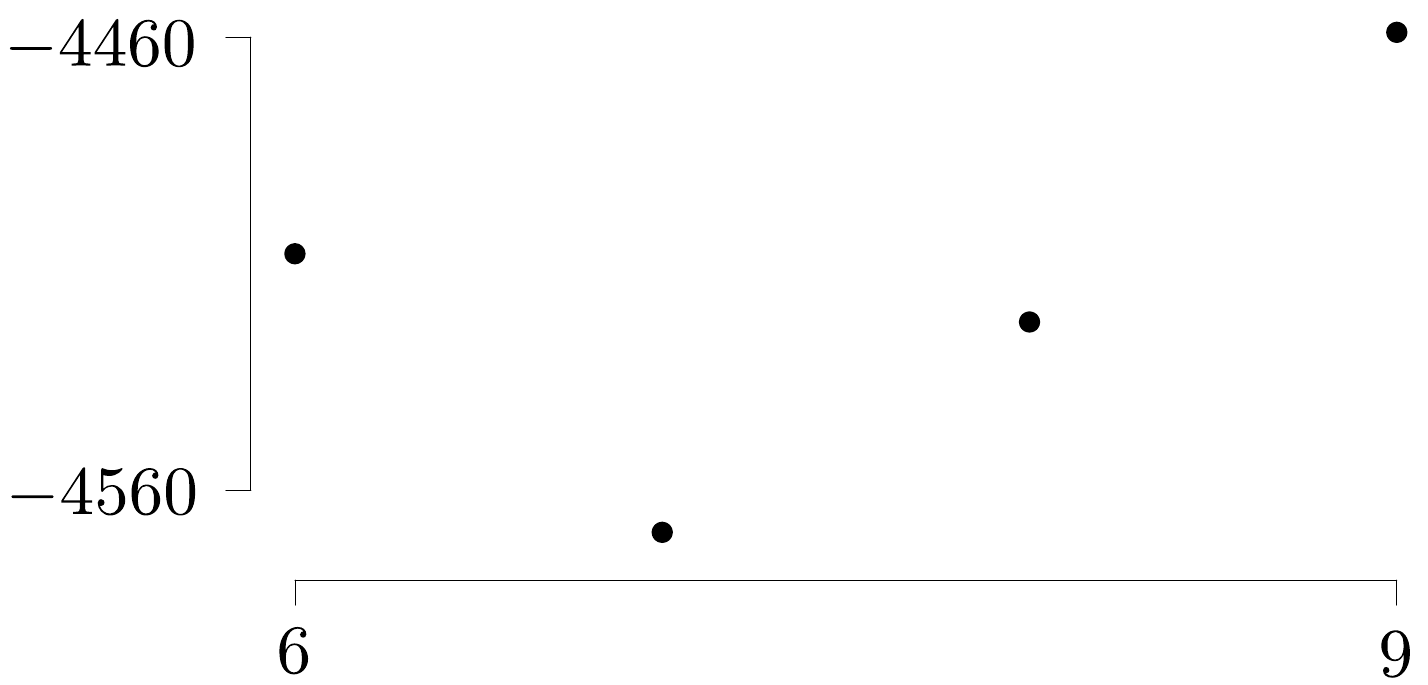}
\caption{\small
	BIC values for four models; the model with the lowest value is selected.
	The values along the vertical axis result from expression~\eqref{eq:bic} whereas the values along the horizontal axis refer to the number of parameters in a model.
}
\label{fig:bic}
\end{figure}

Using expression \eqref{eq:bic}, we calculate the BIC value for each model. 
As illustrated in Figure~\ref{fig:bic}, the least value is obtained for a seven parameter model. 

\section{Discussion}
As illustrated in Figure~\ref{fig:bic}, the best model is composed of seven parameters for which only the middle layer is anisotropic. 
With fewer parameters, the BIC value increases substantially. 
With more parameters, the improvement of the solution is not sufficient to justify  additional parameters. 

Furthermore, we can observe relations between certain parameters. 
In the left-hand plot of Figure~\ref{fig:a0b0,a1chi1}, the values of $a$ and $b$ of the top layer exhibit the correlation coefficient of $-0.96$. 
Consequently, it is impossible to retrieve their individual values, since many of their pairs produce very similar traveltimes. 
The same phenomenon appears if we assume anisotropy in the top layer. 
In the right-hand plot of Figure~\ref{fig:a0b0,a1chi1}, the values of $a$ and $\chi$ in the middle layer exhibit a correlation coefficient of $-0.87$, which makes it impossible to retrieve individual values. 
The same thing occurs with anisotropy in all layers even though the dependency is weaker. 

\begin{figure}
\centering
\includegraphics[scale=1]{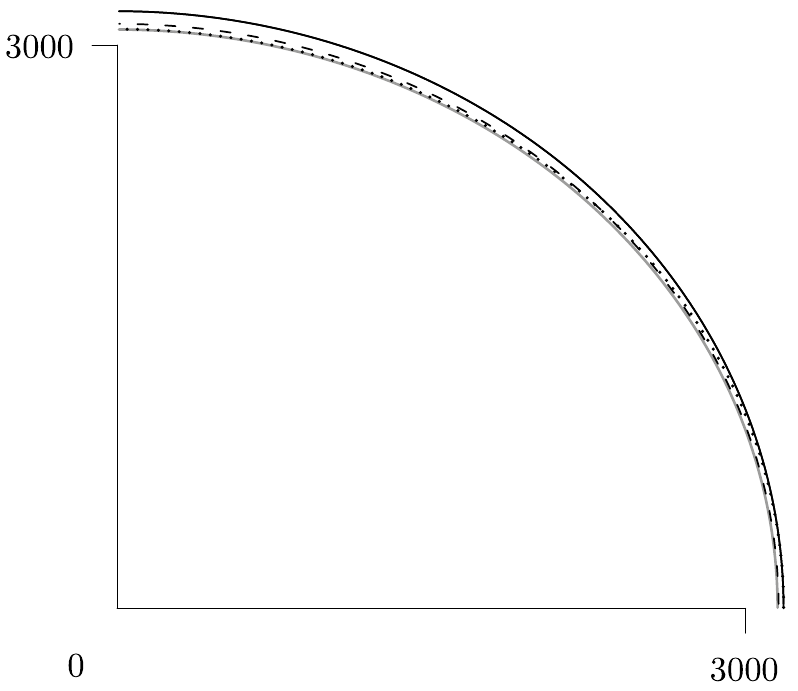}
\caption{\small
	Illustration of ellipses at a depth of 1600\,m depending on the number of model parameters (six: solid black, seven: solid grey, eight: dashed, nine: dotted).
	Since an ellipse has four-fold rotational symmetry, only the first quadrant is illustrated.
	The vertical- and horizontal-velocity components are along the vertical and horizontal axes, respectively; both have units of m/s.
}
\label{fig:elipsy}
\end{figure} 

Anisotropy at specific depths is illustrated by ellipses. 
In Figure~\ref{fig:elipsy}, we show such ellipses for different parameterizations. 
We note that the shapes of these ellipses are modified only slightly by introducing anisotropy in other layers, which is consistent with the choice made based on the BIC criterion.
It supports an inference about the middle layer exhibiting anisotropy, in contrast to $\chi$ being only a fitting---without a physical meaning---parameter for other layers.




Let us comment on our use of BIC in selecting a model parameterization.
In the context of traveltimes alone, a possible parameterization would be a three-layer isotropic model, which would consist of six parameters.
However, we also have external information, which is beyond seismic measurements, namely, the subsurface is comprised of shale.
Hence, we also include anisotropy in our model.  
With the introduction of an extra parameter,~$\chi$\,, BIC suggests that anisotropy should be included in the second layer, within a three-layer model, which results in a seven-parameter model. 
We accept this model for the following reasons.
The isotropy of the first layer can be justified by its position.
The shallowest layer is not subject to significant compaction, which results in preferential alignments and, hence, in anisotropy; the compaction increases with overburden, as a function of depth.
The isotropy of the third layer can be justified---in terms of BIC---by the fact that the resolution of the data decreases with depth. 
Also, since the third layer is the thinnest, the effect of anisotropy on traveltimes is less pronounced than in the middle layer, which contains the anisotropy of the model.

To conclude, let emphasize that, to a large extent, our model is an epistemological analogy to account for observations.
Intrinsically---without external interpretation---there is no ontological claim.
In general, BIC provides a satisfactory model to account for measurements, according to the Bayesian Information Criterion, but not necessarily a model that corresponds to the physical reality.
BIC provides the most empirically adequate model, according to its criteria, even though a more complex model would fit the data better.
This is the very purpose of BIC: to ensure that the model complexity does not surpass the accuracy of data.
\subsection*{Acknowledgements}
The authors wish to acknowledge Elena Patarini for her graphical support and David Dalton for reviewing and editing. 
This research was partially supported by the Natural Sciences and Engineering Research Council of Canada, grant~202259 and AGH University of Science and Technology, Faculty of Geology, Geophysics and Environmental Protection, project~11.11.140.613.
\bibliographystyle{apa}  
\bibliography{DGKSS_arXiv.bib}

\newcommand{\SortNoop}[1]{}
\begin{thebibliography}{}

\bibitem[\protect\astroncite{Aki and Richards}{2002}]{AkiRichards2002}
Aki, K. and Richards, P.~G. (2002).
\newblock {\em Quantitative seismology}.
\newblock University Science Books, 2nd edition.

\bibitem[\protect\astroncite{{\SortNoop{Cerveny}}\v{C}erven\'{y}}{1985}]{cerveny}
{\SortNoop{Cerveny}}\v{C}erven\'{y}, V. (1985).
\newblock The application of ray tracing to the numerical modelling of seismic
  wave fields in complex structures.
\newblock {\em Seismic shear waves}, 15:1--124.

\bibitem[\protect\astroncite{Danek and Slawinski}{2012}]{danek}
Danek, T. and Slawinski, M.~A. (2012).
\newblock Bayesian inversion of {VSP} traveltimes for linear inhomogeneity and
  elliptical anisotropy.
\newblock {\em Geophysics}, 77:R239--R243.

\bibitem[\protect\astroncite{Gierlach and Danek}{2018}]{gierlach}
Gierlach, B. and Danek, T. (2018).
\newblock Inversion of velocity parameters in multilayered elliptical
  anisotropy medium - synthetic data example.
\newblock {\em E3S Web of Conferences}, 66:01017.

\bibitem[\protect\astroncite{Guo et~al.}{2011}]{bic-mt}
Guo, R., Dosso, S.~E., Liu, J., Dettmer, J., and Tong, X. (2011).
\newblock Non-linearity in {Bayesian} {1-D} magnetotelluric inversion.
\newblock {\em Geophysical Journal International}, 185:663--675.

\bibitem[\protect\astroncite{Kaderali}{2009}]{ayiaz}
Kaderali, A. (2009).
\newblock Investigating anisotropy and inhomogeneity using tomographic
  inversion of {VSP} traveltimes: Validation of analytic expressions for
  linearly inhomogeneous elliptically anisotropic models.
\newblock Master's thesis, Memorial University of Newfoundland, St. John's,
  Canada.

\bibitem[\protect\astroncite{Kass and Raftery}{1995}]{kass}
Kass, R. and Raftery, A. (1995).
\newblock Bayes factors.
\newblock {\em Journal of the American Statistical Association}, 90:773--795.

\bibitem[\protect\astroncite{Keller}{1978}]{keller}
Keller, J.~B. (1978).
\newblock Rays, waves and asymptotics.
\newblock {\em Bulletin of the American Mathematical Society}, 84:727--750.

\bibitem[\protect\astroncite{Nelder and Mead}{1965}]{nelder}
Nelder, J.~A. and Mead, R. (1965).
\newblock A simplex method for function minimization.
\newblock {\em The Computer Journal}, 4:308--313.

\bibitem[\protect\astroncite{Priestley}{1982}]{priestley}
Priestley, M.~B. (1982).
\newblock {\em Spectral Analysis and Time Series}.
\newblock Academic Press.

\bibitem[\protect\astroncite{Rogister and Slawinski}{2005}]{rogister}
Rogister, Y. and Slawinski, M. (2005).
\newblock Analytic solution of ray tracing equations for a linearly
  inhomogenous and elliptically anisotropic velocity model.
\newblock {\em Geophysics}, 70:D37--D41.

\bibitem[\protect\astroncite{Schwarz}{1978}]{schwarz}
Schwarz, G. (1978).
\newblock Estimating the dimension of a model.
\newblock {\em Annals of Statistics}, 6:461--464.

\bibitem[\protect\astroncite{Shearer and Chapman}{1988}]{shearer}
Shearer, P.~M. and Chapman, C.~H. (1988).
\newblock Ray tracing in anisotropic media with a linear gradient.
\newblock {\em Geophysical Journal}, 94:575--580.

\bibitem[\protect\astroncite{Slawinski et~al.}{2003}]{slaw03}
Slawinski, M.~A., Lamoureux, M.~P., Slawinski, R.~A., and Brown, R.~J. (2003).
\newblock {VSP} traveltime inversion for anisotropy in a buried layer.
\newblock {\em Geophysical Prospecting}, 51:131--139.

\bibitem[\protect\astroncite{Slawinski and Webster}{1999}]{slaw99}
Slawinski, M.~A. and Webster, P.~S. (1999).
\newblock On generalized ray parameters for vertically inhomogeneous and
  anisotropic media.
\newblock {\em Canadian Journal of Exploration Geophysics}, 35:28--31.

\bibitem[\protect\astroncite{Slawinski et~al.}{2004}]{slaw04}
Slawinski, M.~A., Wheaton, C.~J., and Powojowski, M. (2004).
\newblock {VSP} traveltime inversion for linear inhomogeneity and elliptical
  anisotropy.
\newblock {\em Geophysics}, 69:373--377.

\bibitem[\protect\astroncite{Wang}{2014}]{wang}
Wang, Y. (2014).
\newblock Seismic ray tracing in anisotropic media: A modified {Newton}
  algorithm for solving highly nonlinear systems.
\newblock {\em Geophysics}, 79:T1--T7.

\end{thebibliography}
\end{document}